# Tuning friction at material-nanoparticle-liquid interfaces with an external electric field

B. Acharya[1], C. M. Seed[1], D. W. Brenner[2], A. I. Smirnov[3], J. Krim[1]

[1]Dept. of Physics, North Carolina State University, Raleigh, NC 27695, USA
[2]Dept. of Materials Science and Engineering, North Carolina State University, Raleigh, NC 27695, USA
[3]Dept. of Chemistry, North Carolina State University, Raleigh, NC 27695, USA

*Corresponding author email: jkrim@ncsu.edu

**The use of electrophoretic forces to tune friction at material-nanoparticle-liquid interfaces with static or low frequency (0.6 – 50 mHz) electric fields is reported for the first time. External electric fields were employed to reposition negatively charged $TiO_2$ or positively charged $Al_2O_3$ nanoparticles suspended in water in directions perpendicular to a planar platinum surface of a quartz crystal microbalance, which was then used to monitor frictional shear forces at the interface. Active electro-tunable control of friction has been demonstrated for both $TiO_2$ and $Al_2O_3$ suspensions. For $TiO_2$ suspensions, significant drops in frictional shear forces, not observed for Al2O3, were likely attributed to the presence of molecularly thin interstitial water layers remaining in regions between the $TiO_2$ particles and the substrate. Timescales associated with motion of nanoparticles in directions perpendicular to the surface were also investigated by varying the frequency of the external electric field, and were determined to be similar to those of glass-like or polymeric materials. Overall, the studies reveal that nanoparticles actively driven by electric fields can act as "cantilever-free" atomic force probes capable of "tapping mode" exploration of interfacial properties and nanoscale interactions in geometries inaccessible to optical and micromechanical probes.**

Nanotribology, the study of atomic-scale origins of friction[1], has advanced rapidly in recent years[2] from passive investigations of model systems to active control and rational design of smart lubricant systems[3, 4]. Controlling friction with an external field is of particular current interest[5–13], as it addresses one of the major challenges in the field of tribology: the ability to achieve *in situ* control of friction levels without removing and replacing lubricant materials situated within inaccessible confines of a contact. Recently it has been shown that nanoparticles affect friction levels when added to liquid lubricants[14, 15]. The particles are typically electrically charged to prevent agglomeration[16] and the particle charge has been shown to affect the friction levels. [17] Given the wide range of nanoparticle sizes and functionalized surfaces available[18] and the ability to actively control nanoparticle untethered motion[19-21] with external electric fields, nanoparticles are uniquely suited for active electro-tunable control of friction. Electrophoretic tuning of friction by nanoparticles has not, however, been heretofore demonstrated in either nanoscale or macroscale settings and we are unaware of theoretical assessments the approach's viability. Theoretical modelling of how charged nanoparticles would affect the friction are complicated by hydrodynamic effects that alter diffusion rates near surfaces[22], and the highly dynamic nature of surface charge distributions, which play the major role in friction and adhesion[19, 23]. Local concentrations of nanoparticles, moreover, can be highly heterogeneous in directions normal to the surface, causing abrupt changes in system rheology, jamming and/or increased static friction levels[3, 24–26]. Since wear-induced nanoparticulates are commonplace[27], and since particulates and metal surfaces immersed in liquids are often charged[28], studies of the effects of external electric fields on the tribological response of nanoparticle suspensions are both timely and compelling.

To explore the existence of electro-tunable friction effects in nanoparticulate systems, we employed a Quartz Crystal Microbalance (QCM) immersed in aqueous suspensions of negatively or positively charged $TiO_2$ and $Al_2O_3$ nanoparticles to monitor the friction levels upon applying external electric fields to induce electrophoretic forces[29] for repositioning of the nanoparticles relative to the QCM surfaces (Fig. 1). In addition to inducing electrophoretic motion, electric fields deform the double layers and location of hydrodynamic slip planes, induce dipole moments and impact nanoparticle self-assembly and ordering[19–21]. We hypothesize, therefore that the combined effects will have strong impact on the frictional drag levels impacting the oscillating QCM electrode. The latter method is highly sensitive to particle attachment and slip plane locations [17, 30-36]. Macroscale contacts lubricated by liquids are also influenced by particle attachment and modifications in inter-surface normal forces via changes in the electrical double layer/slip plane location when ratios of applied contact pressure to inter-surface pressure fall below ~10 [37–39]. In cases where the nanoscale QCM response has been directly compared to macroscale tribological response [17,30,31,35], a high degree of correlation has been observed.

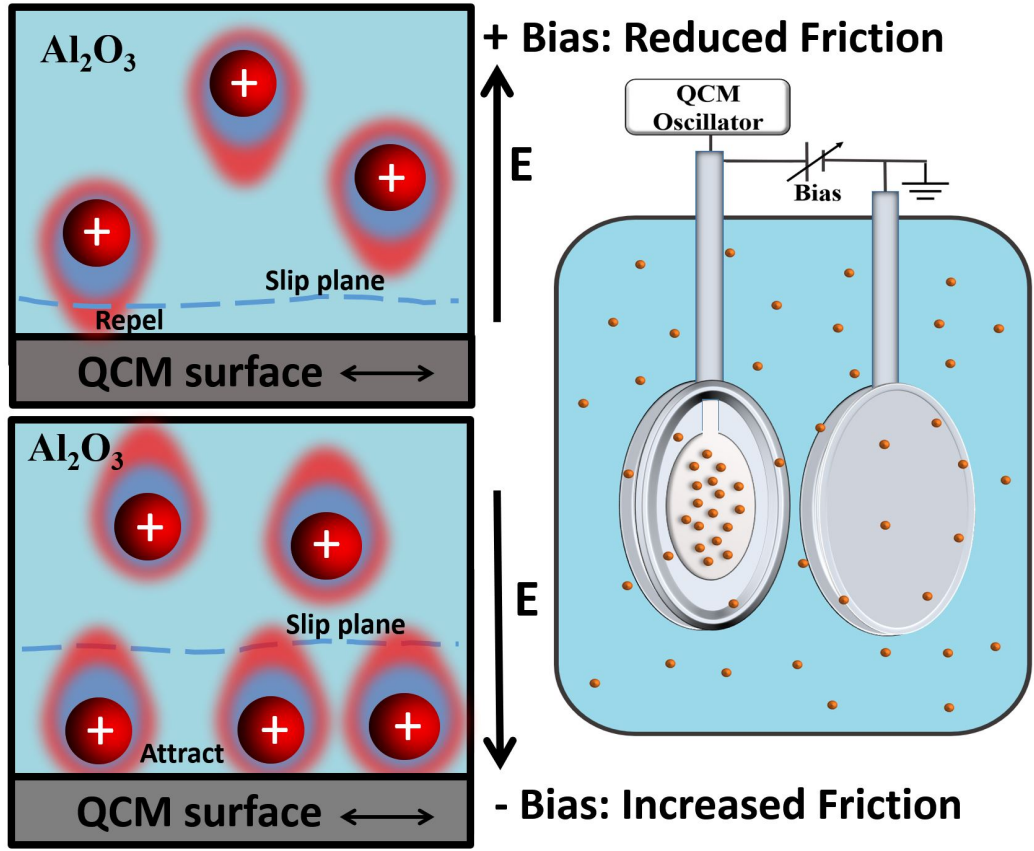

Figure 1. Schematic depiction of an example of the response of positively charged $Al_2O_3$ nanoparticles to positive and negative bias voltages applied to the Pt QCM electrode. Positive (negative) bias voltages repel (attract) the nanoparticles away from (towards) the surface, altering both the hydrodynamic slip plane location and the number of particles in rigid and/or rolling contact with the substrate. Frictional drag levels on the oscillating QCM electrode are highly sensitive to both effects. Friction in macroscale contacts lubricated by liquids are also highly sensitive to the presence of nanoparticles, as well as slip plane locations [37–39].

A QCM's frequency and inverse quality factor $Q^{-1}$ reflect the mass of material dragged along with the oscillatory motion and frictional forces impeding its motion[34, 40]. Attachment of particles on one side of a QCM surface reduces its frequency $f = \omega/2\pi$ by $\delta f = -2.264 \times 10^{-6}(\rho_2 f^2)$ where $\rho_2$ is the mass per unit area of the particles in g/cm$^2$[41]. If the QCM is immersed in a liquid, the additional mass and drag effects are[40]

$$\delta(Q^{-1}) = \frac{2\sqrt{\rho_3 \eta_3 f}}{\sqrt{\pi \rho_q \mu_q}} = 2\alpha; \quad \delta f = -f\frac{\sqrt{\rho_3 \eta_3 f}}{\sqrt{\pi \rho_q \mu_q}} = -f\alpha \tag{1}$$

where $\rho_q$ and $\mu_q$ are respectively the density and the shear modulus of quartz for the AT cut, and $\alpha$ is defined by Eq. (1). A QCM's sensitivity zone is characterized by the shear-wave penetration depth into the liquid given by $\delta = (2\eta_3/\omega\rho_3)^{1/2}$, which for a 5 MHz crystal immersed in water (density $\rho_3 = 1$ g/cm$^3$; viscosity $\eta_3 = 8.9\times10^{-3}$(dyn s/cm$^2$) is 240 nm[34].

5 MHz QCM crystals with Pt electrodes were mounted on a Teflon holder and immersed in DI water, conditions under which platinum develops a negative surface charge density[49]. 40 nm, -33 mV zeta potential TiO$_2$ nanoparticles or 30 nm, +61 mV zeta potential Al$_2$O$_3$ nanoparticles were next dispersed into the solution by increasing their concentration to 1 wt % in six consecutive and equal steps (Fig.2a,b) while monitoring changes in $f$ and the motional resistance $\delta R = 1.16 \times 10^6\ \delta Q^{-1}$. The observed 12-16 Hz negative frequency shift for the Al$_2$O$_3$ particles equates to $\rho_2 = 2 \times 10^{-7}$ g/cm$^2$, or a submonolayer surface coverage with approximate spacing of 150 nm between the particles. The resistance data displayed in Fig. 2b cannot be explained exclusively by changes in the bulk viscosity and density of the suspensions (Eq. 1, blue dots), without invoking changes in slip conditions at the interface[34, 36, 42]. The changes in $f$ and motional resistance $R$ are, therefore, collectively attributable to changes in suspension density and surface effects associated with particle uptake, interfacial slippage and changes in the slip plane location.

Figures 2a,b show that resistance to drag forces *decreases* when TiO$_2$ particles are added, reflecting an increase in slip length[34, 42]. This is consistent with reduced friction levels reported for QCM's with gold and stainless steel electrodes (which, like Pt develop negative surface charge in water) immersed in negatively charged suspensions and also macroscale systems lubricated by them[17, 30, 32, 36]. The data are consistent with Al$_2$O$_3$ particles residing directly on the electrode surface and TiO$_2$ particles residing atop of an interstitial water layer [17,32]. The reduced friction levels for the TiO2 suspensions are also reminiscent of the phenomenon of hydration lubrication [33] whereby the water layers surrounding charges are difficult to remove from the enclosed charge, but upon shear exhibit a high level of fluidity.

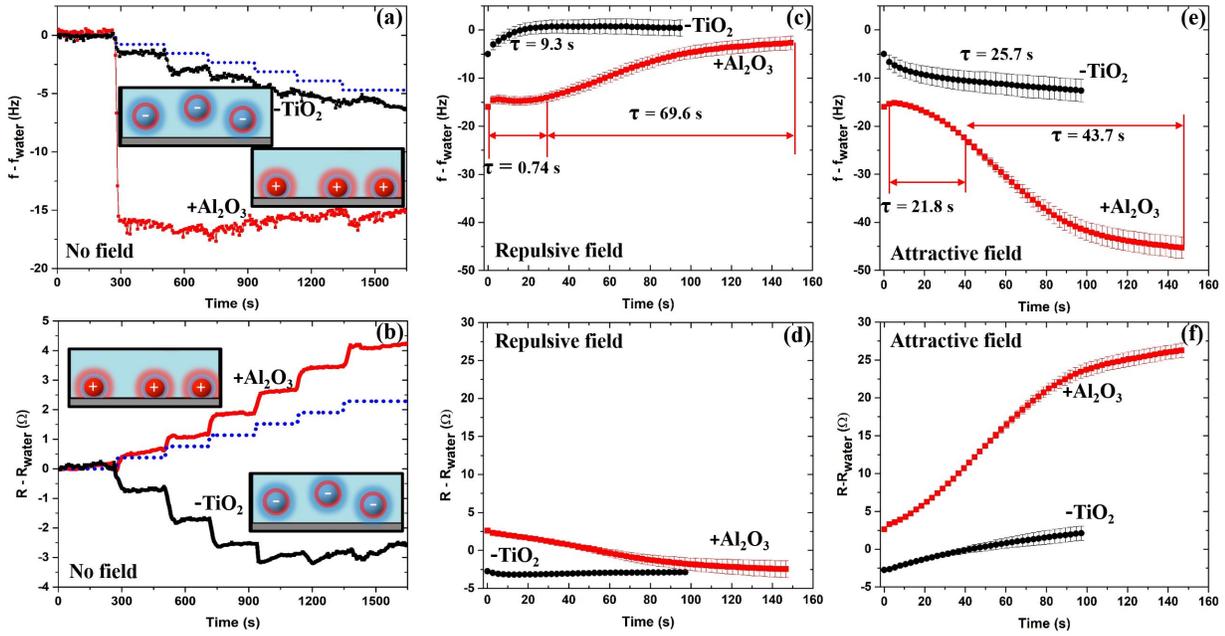

**Figure 2.** QCM frequency (a,c,e) and resistance (b,d,f) versus time for three static field conditions. (1) Zero field (a,b) as the concentration of positively (negatively) charged $Al_2O_3$ ($TiO_2$) nanoparticles is increased from 0 to 1 wt% in six equal increments. The data are explained by a combination of increasing suspension density (Eq.1, blue dotted lines), particle uptake, and changes in frictional drag forces. (2) Repulsive field conditions (+100 N/C bias for $Al_2O_3$ and -100 N/C for $TiO_2$) (c,d) for 0.67 wt% suspensions, and (3) Attractive field conditions (e,f) achieved by reversing the field directions. (b). Repulsive (attractive) electric fields give rise to reductions (increases) in friction and removal (addition) of nanoparticles within the QCM's sensitivity zone over characteristic time periods of t = 1-100 s (c-f). The two time constants for $Al_2O_3$ particles are attributed to particles residing directly on the surface and particles in suspension. $TiO_2$ nanoparticles notably reduce frictional drag forces (R-$R_{water}$) while $Al_2O_3$ nanoparticles increase them, consistent with the $Al_2O_3$ particles residing directly on the electrode surface and $TiO_2$ particles residing atop of an interstitial water layer (see text .[17,32]

The QCM electrodes were next immersed in 0.67 wt% suspensions of either $TiO_2$ or $Al_2O_3$ nanoparticles and situated 1.5 cm away from a second, identical electrode which was grounded (Fig. 1). Static or sinusoidal bias potentials with frequencies ranging from 0.6 to 50 mHz were then applied to the QCM sensing electrode. The applied potential of ± 1.5 V (peak or constant) produced an electric field of ± 100 N/C in the region between the QCM sensor and the grounded electrode. The experiments were repeated at least 5 times each for the case of static electric fields and at least 5 cycles for each of the AC electric field frequencies.

Figures 2c,d and 2e,f respectively, display QCM response to repulsive and attractive external fields for 0.67 wt% suspensions and provide definitive evidence that an electric field can be employed to actively reposition particles to tune the friction levels. For $TiO_2$ particles, the maximum drop in motional resistance occurs under slightly repulsive conditions, and the particles move towards the surface ($\tau$ = 25.7 s) more slowly than they do away from the surface ($\tau$ = 9.3 s). This is consistent with the known phenomenon of slowing Brownian motion of the nanoparticles near the surfaces[43–45]. Assigning 10 s as the time to transit the penetration depth of 240nm, the estimated average outbound particle speed is v = 24 nm/s.

The $Al_2O_3$ suspension response occurs at two distinct time scales (Fig.2c,e), attributed to distinct transient behaviors of the particles in a suspension and those in a bound surface layer. Similar to the $TiO_2$ particles, the time constant attributable to $Al_2O_3$ particles in suspension is longer for the approaching particles ($\tau$ = 21.8s) than for the receding ones ($\tau$ = 0.74 s). This is consistent with a bound layer being present on the surface that is slowly removed by the repulsive electric field, with a removal time constant (t = 69.6s) longer that to reform on the surface under an attractive field ($\tau$= 43.7s)

Figure 3 compares the QCM response to the AC electric field for three frequencies (see the supplemental information for the full set). Since the removal times associated with the surface-bound $Al_2O_3$ particles are close to 70 s for a constant field of 100 N/C and presumably slower for lesser fields, the removal and reattachment of the particles are effectively frozen out. This is evident in Fig. 3a,b where the frequency amplitude for $Al_2O_3$ remains close to zero for the repulsive half-cycle, with no evidence of the particle removal. The data for repulsive potential therefore reflect $Al_2O_3$ particles in suspension since the particles on the surface require longer time periods to respond and their motion is effectively frozen out at these timescale.

QCM data for $TiO_2$ nanoparticles exhibit a behavior consistent with a driven damped motion albeit asymmetrically in the directions approaching and receding from the surface. As the frequency of the external electric field increases, the system response increasingly lags in phase behind the drive frequency. The phase angle lag increases with the drive frequency while the amplitude of the system response diminishes (Fig.3,4). The data are not however simply sinusoidal in nature: the double spike-like features close to +1 V appears during the attraction half-cycle and are observed for frequencies lower than 0.06 rad/sec. The trailing spikes of this feature disappear for the field frequencies 0.06 rad/sec and higher. We note that no spikes of any kind were observed under static electric fields (cf. Fig. 2). In addition the smaller response for the repulsive regime for the $Al_2O_3$ particles than the $TiO_2$ particles in Fig. 4 is likely attributable to the fact that the data reflect the $Al_2O_3$ particles in suspension, which are farther from the surface than the $TiO_2$ particles because the bound layer is present.

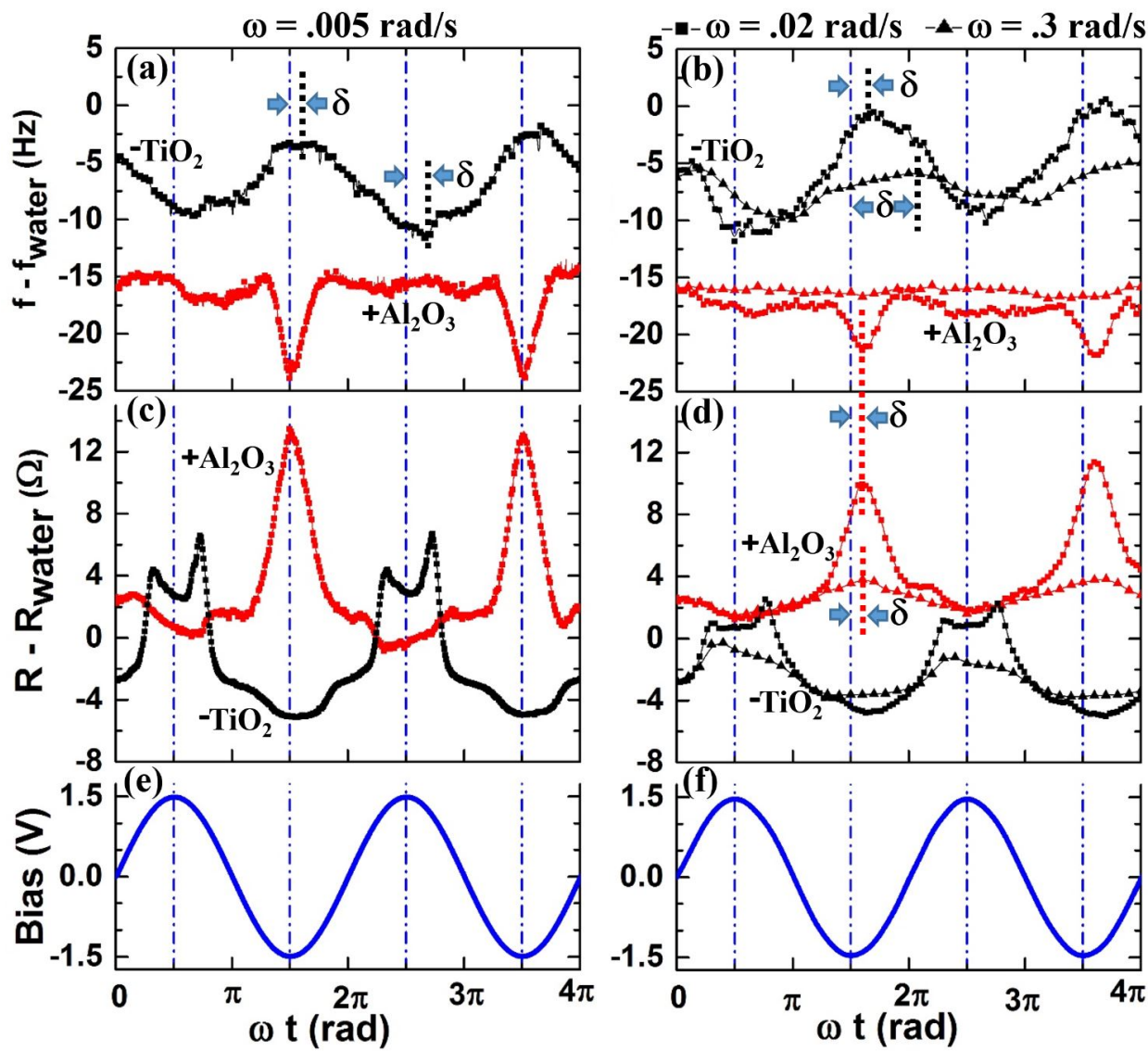

Figure 3. QCM frequency (a,b,) and resistance (c,d) response versus ωt for 0.67 wt% suspensions of $TiO_2$ and $Al_2O_3$ nanoparticles exposed to AC electric fields (100 N/C peak amplitude) oscillating at 0.005 rad/s (a,c), 0.02 rad/s (b,d, squares) and 0.3 rad/s (b,d triangles). Shifts in opposite directions for a given bias voltage are attributed to the nanoparticles' opposite charges. Strong asymmetry between the attractive and repulsive biases for the $Al_2O_3$ data is consistent with the particles residing on or near a hard wall substrate (a,c). The shape of the response for the $TiO_2$ data may arise from reorientation of interfacial water molecules [13, 49], which are not present for $Al_2O_3$ particles. At higher frequencies (b,d) the response diminishes and the phase lag δ increases [46].

The spikes are not attributable to electrolysis as no bubbles were observed either visually or in a form of the characteristic QCM response[47]. It is known, however, that the voltage levels corresponding to the electrolysis are closely linked to the thickness of the oxide layers on the platinum surface[27, 48–50]. The latter may significantly increase the voltage threshold at which the steady-state electrolysis first occurs[27, 48–50]. The QCM response features could be associated with field-induced reorientation of water molecules at the surface[13, 51], reorientations of the nanoparticles themselves or non-monotonic charging of the Pt electrode as it passes through points of zero charge density[13,49]. For example, the QCM data for for $TiO_2$ suspensions can be explained by a reorientation of the interstitial water molecules. A strong interaction between platinum and the hydrogen atom of the water molecules being oriented towards the surface has been discussed in the literature[48]. A positive electric field could possibly reverse this interaction by changing the platinum electronic configuration but a negative field would not. Indeed, the effect is prominent for the positive electric fields applied to $TiO_2$ and small features are in fact also observed in the $Al_2O_3$ data that could be attributable to the water reorientation. No such features are present for the negative fields. The spikes in the resistance are consistent with changes in the interfacial friction levels, which are highly sensitive to physical orientations of molecules at interfaces. The $Al_2O_3$ data in Figs. 3 and 4 also display additional features that may be consistent with the primary and secondary potential minima as the particles are pressed closer towards the electrode. We also note that non-monotonic charging of the Pt electrode as it passes through points of zero charge density can also explain the observed phenomena. Overall, while the origin of this interesting phenomenon is not fully clear at this point, the data reveal nanoscale interactions in geometries inaccessible to optical and micromechanical probes[52, 53]. The data presented herein reflect the *collective* response of the nanoparticles to an external electric field directed perpendicular to the electrode surfaces. Nanoscale particles in general exhibit Brownian motion due to frequent and random collisions with the solvent molecules. In bulk the average displacement is zero, because the collisions occur randomly in all directions. A characteristic diffusion time $\tau_D = a^2/6D = (\pi\eta_3 a^3)/(k_B T)$ is then defined as the average time required for a spherical particle to diffuse over a distance equal to its own radius[54, 55], where a is the particle radius and D is the diffusion coefficient. For the systems studies here, $\tau_D = 3 \times 10^{-6}$ s and $7.7 \times 10^{-6}$ s for $TiO_2$ and $Al_2O_3$ nanoparticles respectively, significantly shorter than the collective particle rearrangement time scales measured in our experiments.

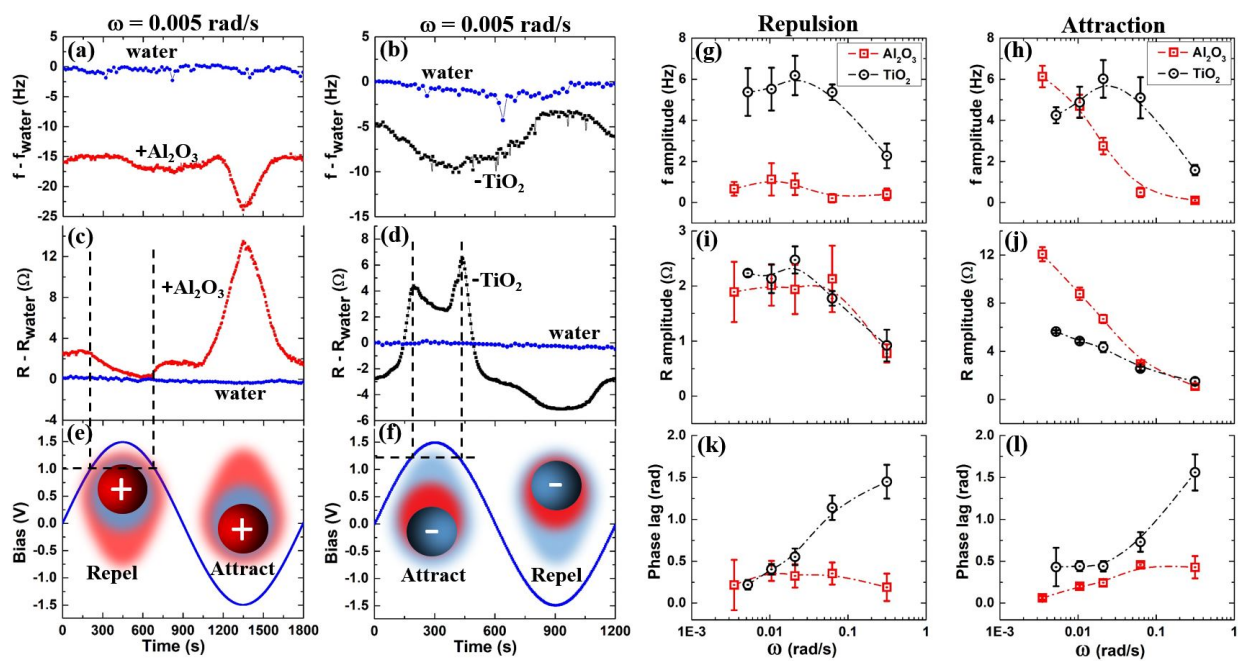

**Figure 4.** QCM frequency (a,b) and resistance (c,d) response for 0.67 wt% suspensions of $Al_2O_3$ and $TiO_2$ nanoparticles and pure water (blue) for one cycle at 0.005 rad/s (e,f), and as a function of ω (g-l), including phase lag (k,l), for the repulsive (g,i,k) and attractive (h,j,l) regimes of the applied bias voltage. The features denoted by dashed lines (c-f) are close to +1V and may indicated reorientation of interfacial water molecules at glass-like time scales and/or the Pt surface passing through points of zero charge density[13, 49].

It is estimated that $TiO_2$ particles transit the penetration depth of 240 nm in about 10 s, at an estimated speed of v =24 nm/s. This is substantially lower than a typical electrophoretic speed of 2,500 nm/s [28, 56, 57]. For a force of qE = (-2.8 × $10^{-17}$C)×(-100 N/C)= 2.8 ×$10^{-15}$ N, this corresponds to an electrophoretic mobility value µ = v/E = (24x$10^{-9}$m/s)/(100N/C) = 2.4 ×$10^{-10}$ $m^2$/Vs, which is also substantially lower than the typical theoretical values of ca. $\epsilon\zeta/\eta$ = (6.95 x $10^{-10}$)(0.0327) /(8.9 ×$10^{-4}$) = 2.6 ×$10^{-8}$ $m^2$/Vs. The average applied force is, however, a sum of both the force caused by the externally applied electric field and the attractive substrate forces once the particle exits the potential minimum. The method reported here then provides a unique means to probe the interfacial forces in systems that may not be accessible to optical or micromechanical methods[34, 52, 53].

For $TiO_2$ suspensions, the phase angle data can be compared to other systems in terms of the governing time scales. For example, the angular frequency at which the $TiO_2$ system response lags the drive voltage by 45° = 0.785 rad is ω = 0.03-0.06 rad/s, which corresponds to 0.005 – 0.01 Hz. In contrast, numerical studies of colloidal particles in aqueous suspensions with similar zeta potentials report phase shifts of 45º near 1000 Hz[56, 57] while electrochemical capacitors comprised of electrical double layers fall in the range 0.1 – 1000 Hz[58]. The present values are similar in magnitude to those observed in glassy and gel-like systems as well as materials confined in pores[54, 59, 60]. Slow dynamics have also been observed for water molecules in confined geometries[62]. In unconfined geometries the first layer of water at a metal water interface has also been reported to exhibit an ice-like structure under positive bias close to the point of zero charge[51].

It is important to distinguish the timescales associated with the QCM response to repositioning of of the nanoparticles by electrophoretic forces in the directions perpendicular to the surface from the response of the system to the shear motion of the electrode of the QCM. For the case of shear motion, the reduced friction levels for the $TiO_2$ suspensions are reminiscent of the phenomenon of hydration lubrication [33, 63-65] whereby the water layers surrounding charges are difficult to remove from the enclosed charge, but upon shear exhibit a high level of fluidity. This phenomenon occurs when water that surrounds charges is trapped between surfaces (for example, the nanoparticle and the substrate) and becomes highly resistant to being squeezed out while remaining quite fluid under shear. In contrast, the water layers between $Al_2O_3$ are readily squeezed out, and comparable lubricity is not observed. The relative contributions of the various effects constitutes an open challenge for modeling studies[61], and the results presented here open a new venue for controlled studies of these individual contributions.

In summary, we have applied external electric fields to induce electrophoretic forces for repositioning of charged nanoparticles relative to QCM surfaces to demonstrate an active electro-tuning of the interfacial friction levels for the first time. Field-induced movements of particles away from the interfaces tended to reduce friction, while pushing the nanoparticles into the surfaces tended to increase the friction levels. Our study reveals that the charged nanoparticles manipulated by external electric fields can act as "cantilever-free" atomic force probes capable of a "tapping mode" exploration of interfacial properties and nanoscale interactions in geometries inaccessible to optical and micromechanical probes. We moreover point the way to a wealth of readily attainable information in future investigations employing electric fields with increased complexity.

This work was supported by the US National Science Foundation, Materials Genome Initiative Project #DMR1535082. We are grateful to J. Batteas, S. Kim, J. Kelin, and I. Szlufarska for useful discussions. A. Marek and V. Perelygin are respectively thanked for assistance with recording the pH and zeta potentials of the samples.